\documentclass[aps,pre,twocolumn,showpacs]{revtex4}
\usepackage{graphicx,epsfig}
\begin{document}
\title{Phase separation in coupled chaotic maps on fractal networks}
\author{K. Tucci$^1$, M. G. Cosenza$^2$, and O. Alvarez-Llamoza$^3$}
\affiliation{$^1$SUMA-CESIMO,
Universidad de Los Andes, M\'erida, Venezuela\\
$^2$Centro de Astrof\'{\i}sica Te\'orica, Facultad de Ciencias,
Universidad de Los Andes, \\ Apartado Postal 26 La~Hechicera,
M\'erida~5251,
Venezuela\\
$^3$Departamento de F\'{\i}sica, FACYT, Universidad de Carabobo,
Valencia, Venezuela}
\date{\today}
\begin{abstract}
The phase ordering dynamics of coupled chaotic maps on fractal
networks are investigated. The statistical properties of the systems
are characterized by means of the persistence probability of equivalent
spin variables that define the phases. 
The persistence saturates and phase domains freeze for all values of the
coupling parameter as a consequence of the 
fractal structure of the networks, in contrast to the phase transition behavior
previously observed in regular Euclidean lattices. 
Several discontinuities and other features found in the
saturation persistence curve as a function of the coupling 
are explained in terms of changes of stability of local 
phase configurations on the fractals.
\end{abstract}
\maketitle

Coupled map lattices have provided fruitful and computationally
efficient models for the study of a variety of dynamical
processes in spatially distributed systems \cite{Chaos}. The
discrete-space character of coupled map systems makes them specially
appropriate for the investigation of spatiotemporal dynamics on
nonuniform or complex networks. Phenomena such as pattern
formation, spatiotemporal intermittency, nontrivial collective
behavior, synchronization, etc., have been extensively studied in
coupled map systems defined on fractal lattices \cite{CK2}, 
hierarchical structures \cite{Gade}, trees \cite{Tree}, 
 random graphs \cite{Vol}, small-world networks \cite{Kay}, 
and scale-free networks \cite{Amri}.

Recently, there has been much interest in the study of
the phase-ordering properties of systems of coupled chaotic maps
and their relationship with Ising models in statistical physics
\cite{Chate,Chate2,Wei,Stra,Just}. These works have invariably assumed 
the phase competition dynamics taking place on a uniform
Euclidean space; however, in many physical situations
the medium that supports the dynamics can be nonuniform on some length scales.
The nonuniformity may be due to the intrinsic heterogeneous nature of 
the substratum such as  porous or fractured media, or it may arise from
random fluctuations in the medium.
This article investigates the phenomenon of phase ordering in 
coupled chaotic maps on fractal networks as a model for studying this 
phenomenon on nonuniform media. The class of fractal
networks being considered corresponds to generalized Sierpinski gaskets (GSG)
embedded in Euclidean spaces of arbitrary dimension $d$ \cite{CK2}.
In particular, this model of coupled maps on fractal networks yields a
situation to explore the role that the connectivity of the
underlying lattice plays on the statistical properties of
phase ordering processes in nonlinear coupled systems.

Deterministic fractal networks, such as GSG, can be generated in
any \mbox{$d$-dimensional} Euclidean space as follows \cite{CK2}. At the
\mbox{$n$th} level of construction, the fractal consists of
$N=(d+1)^{n}$ \mbox{$d$-dimensional} hypertetrahedral cells whose
coordinates can be specified by a sequence
$(\alpha_{1}\alpha_{2}\ldots\alpha_{n})$, where $\alpha_m$ can
take any value in a set of $(d+1)$ different symbols which can be
chosen to be $\{0,1,\ldots,d\}$. At level $n+1$, each cell
$(\alpha_{1}\alpha_{2}\ldots\alpha_{n})$ splits into $d+1$ cells
scaled down by a longitudinal factor of two, and which are now
labeled by $(\alpha_{1}\alpha_{2}\ldots\alpha_{n}\alpha_{n+1})$,
where the first $n$ symbols of the sequence
are the same as the parent cell.
Given this construction rule, the fractal dimension of
the GSG is $d_f=\log(d+1)/\log 2$.
A label
$(\alpha_{1}\alpha_{2}\ldots\alpha_{n})$ can be written as
$(\alpha_{1}\alpha_{2}\ldots\alpha_{n-s}\alpha_{n-s+1}^{s})$ for
some $s \in \{1,2,\ldots,n\}$, where $\alpha_{i}^{s}$ means the
sequence of $s$ identical symbols $\alpha_{i}$. The cell with this
label has a neighborhood ${\cal
N}_{(\alpha_{1}\alpha_{2}\ldots\alpha_{n})}$ with $d+1$ elements
labeled by $(\alpha_{1}\ldots\alpha_{n-1}(\alpha_{n}+j))$
$(j=1,2,\ldots,d)$ and
$(\alpha_{1}\ldots\alpha_{n-s+1}\alpha_{n-s}^{s})$, where the
addition $\alpha_{i}+j$ is defined modulo $(d+1)$. 
If $s=n$, then the cell is one of the $d+1$
vertices of the gasket, labeled by $(\alpha_{1}^{n})$, and it has
only $d$ neighbors belonging to the same parent cell. An integer
index $i=0,1,\ldots,(d+1)^{n}-1$, can be assigned to each cell of
the lattice at the level of construction $n$ by the rule
$i=\sum_{m=1}^n\alpha_m (d+1)^{n-m}$.

The equations describing the dynamics of the diffusively coupled map
system defined on these fractal networks at level of construction $n$, embedded
in a \mbox{$d$-dimensional} Euclidean space, are 
\begin{equation}
\label{ec1}
\begin{array}{ll}
x_{t+1}(\alpha_1\ldots\alpha_n)=
(1-\epsilon)f\left(x_t(\alpha_1\ldots\alpha_n)\right)+  &  \\
      &   \\
  \frac{\epsilon}{d+1} \,
    \sum_{ (\beta_1\ldots\beta_n) \in
{\cal N}_{(\alpha_1\ldots\alpha_n)}}
f\left(x_t(\beta_1\ldots\beta_n)\right),
\end{array}
\end{equation}
where $x_t(\alpha_1\ldots\alpha_n)$ gives the state of the cell
$(\alpha_1\ldots\alpha_n)$ at discrete time $t$; $(\alpha_1\ldots\alpha_n)$
and $(\beta_1\ldots\beta_n)$ label the $(d+1)^n$ cells on the gasket;
$\epsilon$ is a parameter measuring the coupling strength between neighboring
sites, and
$f(x)$ is a nonlinear function that expresses the local dynamics.
Equation~(\ref{ec1}) also applies to the $(d+1)$ vertex cells of the
fractal network, except that the coefficient of the sum is 
$\epsilon/d$, since each of these cells have $d$ neighbors.

As local dynamics, we assume a piecewise
linear, odd  map \cite{Chate}
\begin{equation}
f(x)= \left\{
\begin{array}{ll}
-2\mu/3-\mu x & \mbox{if $x \in [-1,-1/3]$}\\
\mu x         & \mbox{if $x \in [-1/3,1/3]$}\\
2 \mu/3 - \mu x & \mbox{if $x \in [1/3, 1]$} .
\end{array}
\right.
\end{equation}
When the parameter $\mu \in [1,2]$, the map possesses
two symmetric  chaotic attractors contained in the intervals
$I^{\pm} = [\pm \mu (2- \mu )/3, \pm \mu/3]$, and separated by a
gap $I^o = [ - \mu (2- \mu )/3, \mu (2- \mu )/3 ]$.
For values of $\mu$  close to two, the size of the chaotic intervals
is larger than the gap. Then the local states have two well defined symmetric
phases that can be characterized by spin variables defined
as the sign of the state at time $t$,
$\sigma_t(\alpha_1\ldots\alpha_n)=\mbox{sign}(x_t(\alpha_1\ldots\alpha_n))$.

To study the phase-ordering phenomenon of the coupled maps
on fractal networks we fix the local map parameter at the value $\mu=1.9$ and
set the initial condition as follows:
if the number of cells $(d+1)^{n}$ in a lattice is even ($d$ odd),
exactly one half of the sites
are randomly chosen and assigned random values uniformly distributed on
the interval $I^+$, while the other half are similarly assigned values on
$I^-$. If the number of cells in a lattice is odd ($d$ even), then the
state of the remaining cell is assigned at random on either interval
$I^+$ or $I^-$.

The statistical properties of the phase-ordering process on the
fractal networks can be characterized
by using the persistence probability $P_t$, defined as the fraction of
cells which have not changed sign up to time $t$ \cite{Derrida}.
Figure~1 shows $P_t$ as a function of time for the
GSG embedded in $d=3$, for several values of the
coupling parameter.
For some ranges of the coupling, the persistence saturates in a few
iterations, while for some other ranges of $\epsilon$, $P_t$ reaches
its saturation value more slowly.
In contrast, in regular Euclidean lattices the persistence saturates
for small couplings, while it decays
algebraically in time for coupling strengths greater than some critical value
\cite{Chate}.
\begin{figure}[h]
\centerline{\hbox{
\epsfig{file=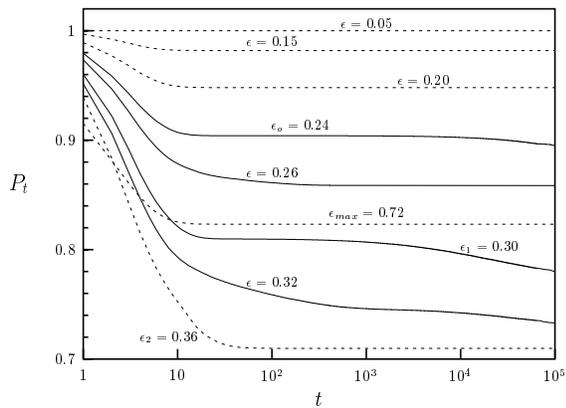,width=0.35\textwidth,clip=,angle=90}
}}
\caption{Persistence probability as a function of time for the
Sierpinski gasket embedded in Euclidean dimension $d=3$ for different values of
the coupling $\epsilon$. Lattice size is $N=4^9$.
Dotted curves correspond to values of $\epsilon$ for which $P_t$ reaches
its saturation value quickly.}
\end{figure}

The domains formed by the two phases on fractal lattices
reach a frozen configuration for all values of the coupling
$\epsilon \in [0,1]$.
Figure~2 shows the asymptotic patterns of the phase separation process on a
GSG embedded in $d=3$ at level of construction $n=3$, for
several values of the coupling. Note that the configuration of
the blocked phase domains changes as the coupling is varied.
The domain configurations can be characterized by the
fraction of sites in a given phase that have $k$ neighbors in that same
phase at time $t$,
denoted by $F_t(k)$, with $k=0,1,\dots,d+1$.
For example, consider the pattern displayed in Fig. 2(a) where
there are several sites in one phase,
as those indicated by arrows, having all of their four neighbors in
the opposite phase. Thus the asymptotic fraction $F_\infty(0)$
is greater than zero in this case.
In Fig. 2(b), as $\epsilon$ increases, $F_\infty(0)$ becomes zero, but
$F_\infty(1)$ is finite since there are sites in one phase,
as those signaled by arrows, having just one neighbor in that same phase.
\begin{figure}[h]
\centerline{\hbox{
\epsfig{file=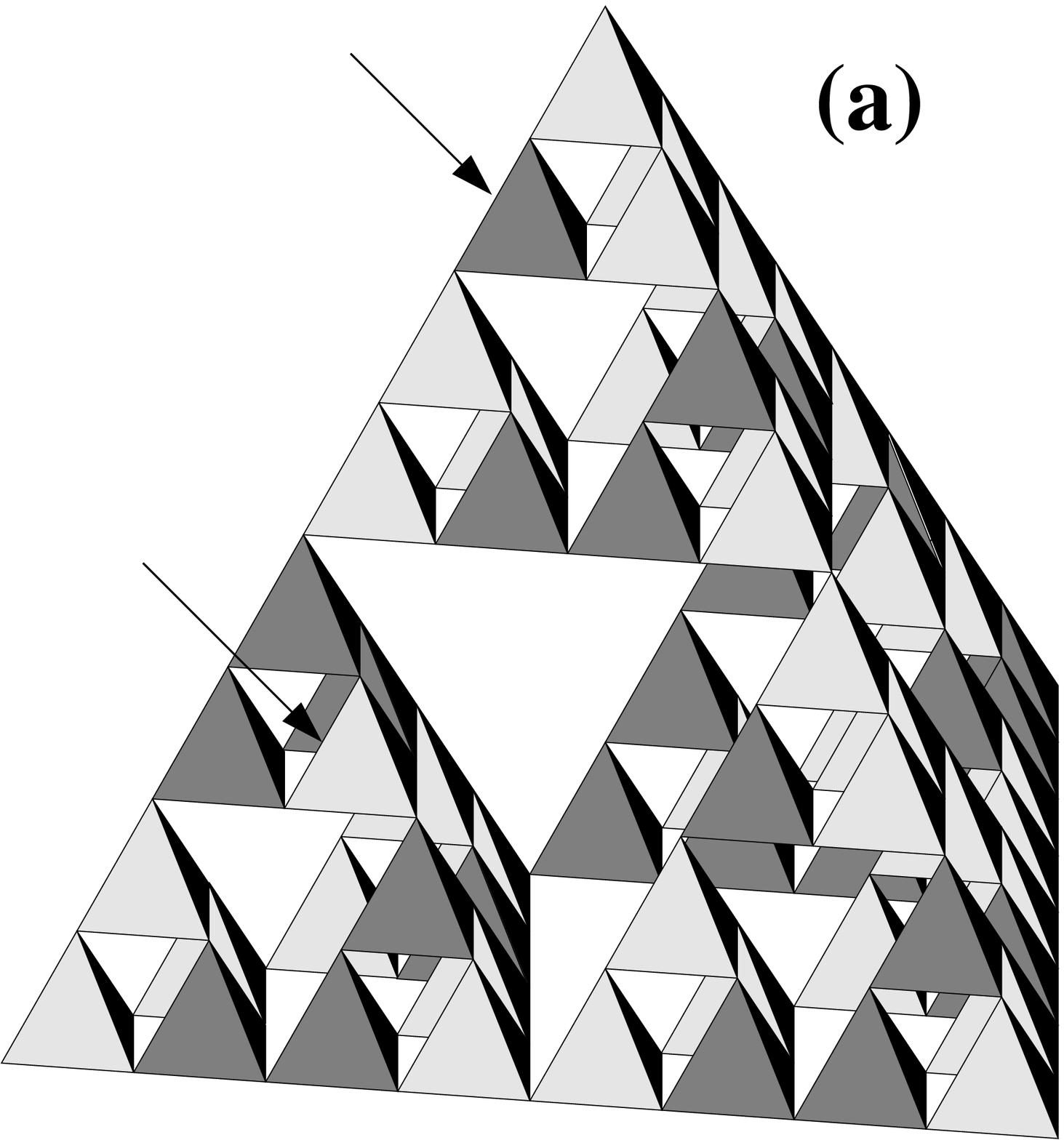,width=0.20\textwidth,clip=,angle=0} \qquad
\epsfig{file=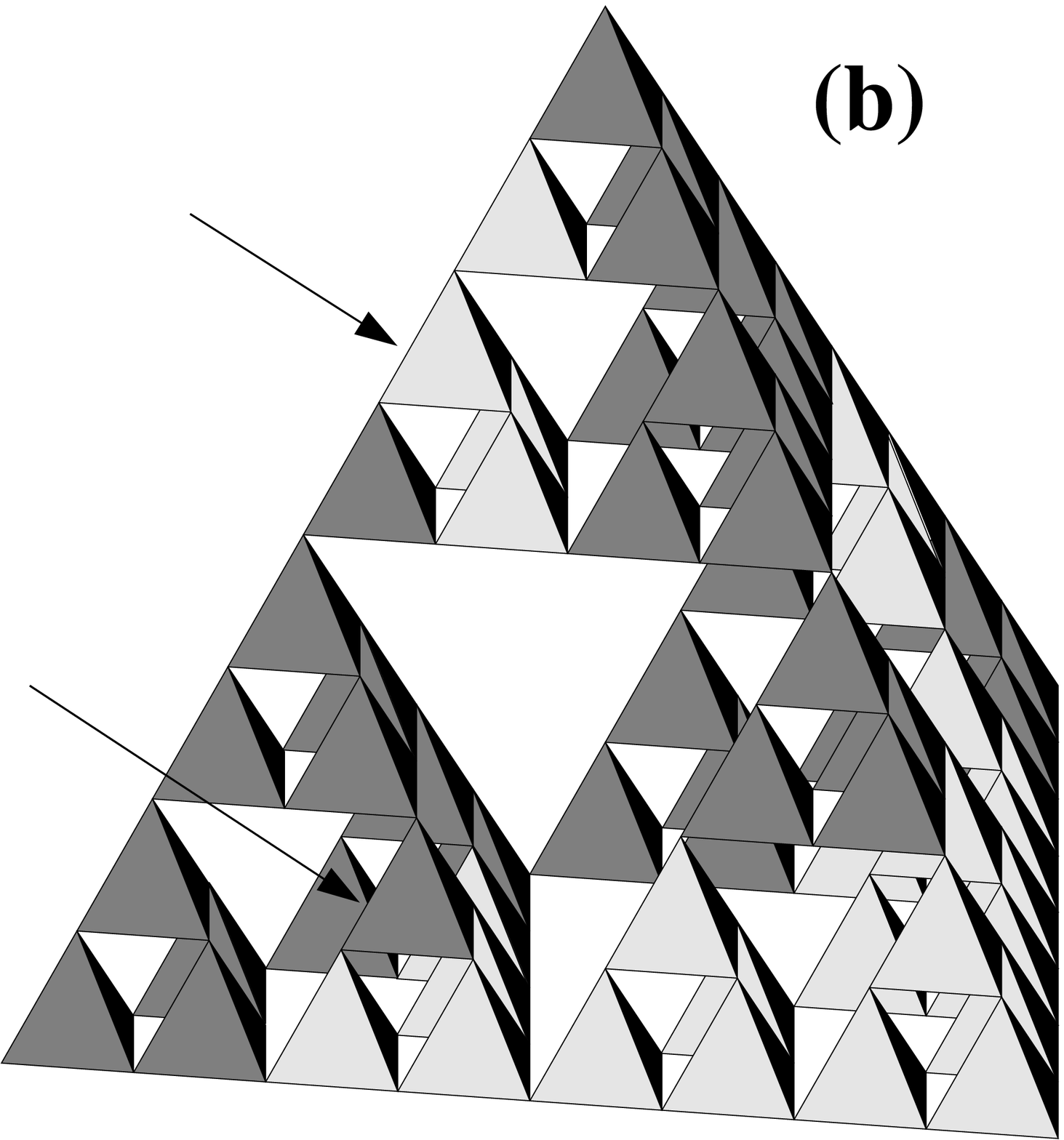,width=0.20\textwidth,clip=,angle=0}
}}
\ \\ \
\centerline{\hbox{
\epsfig{file=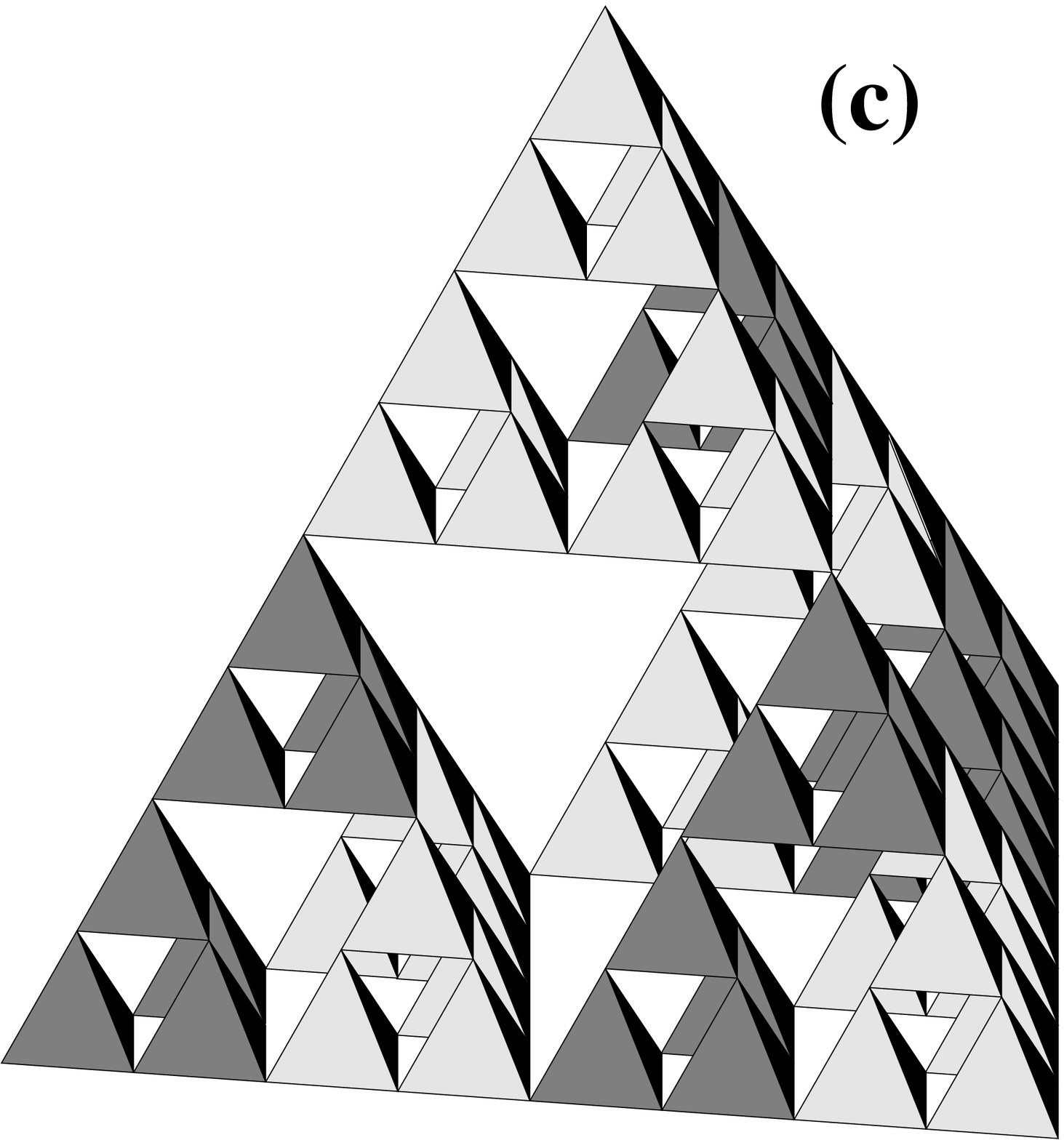,width=0.20\textwidth,clip=,angle=0} \qquad
\epsfig{file=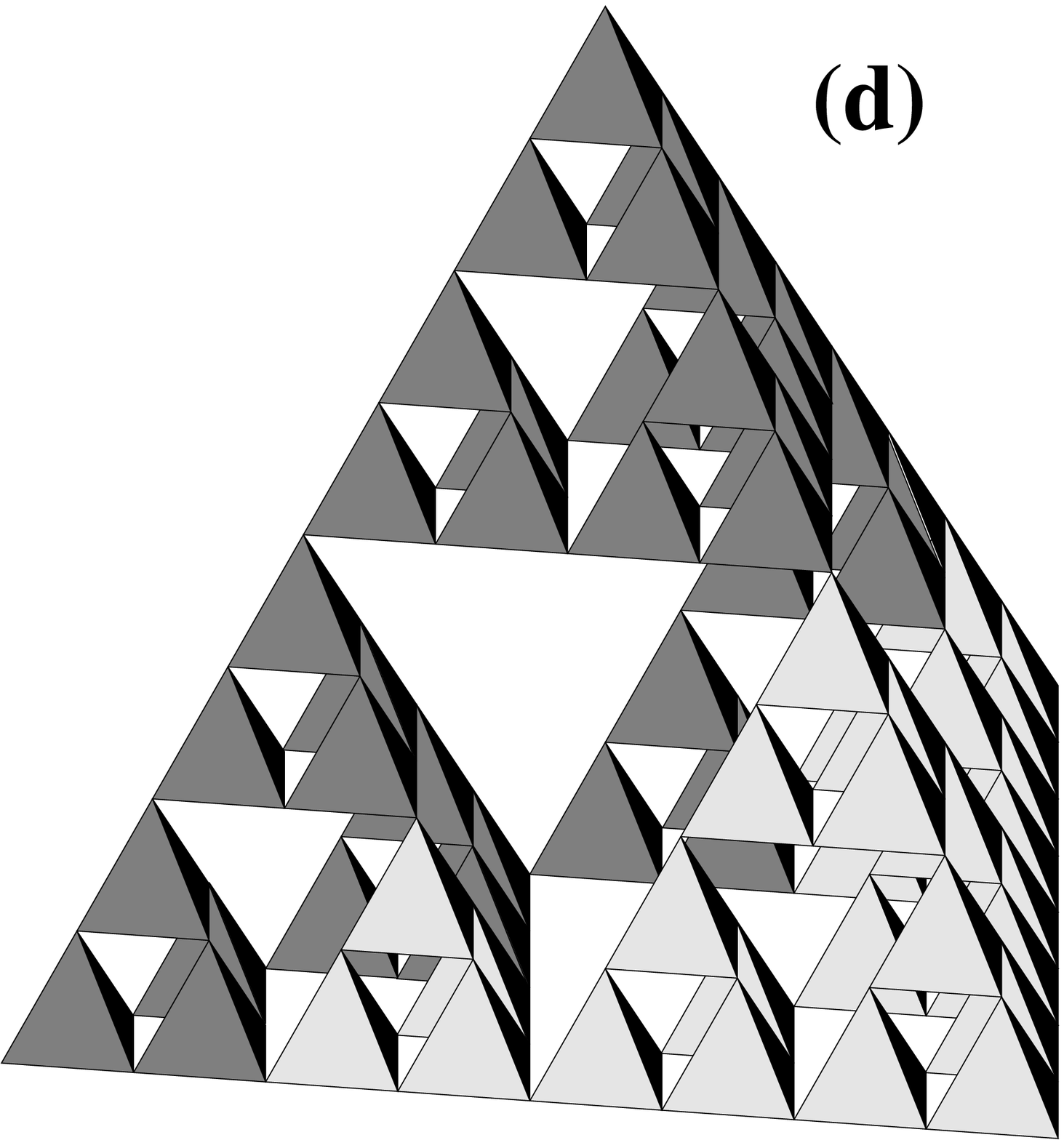,width=0.20\textwidth,clip=,angle=0}
}}
\caption{Phase separation on the GSG  embedded
in Euclidean dimension $d=3$ at level of construction $n=3$, for
several values of the coupling. Dark (light) color corresponds to the
positive (negative) phase. Arrows signal local configurations described
in the text.
(a) $\epsilon=0.20$; (b) $\epsilon=0.26$;
(c) $\epsilon=0.36$; (d) $\epsilon=0.72$.}
\end{figure}

The relationship between the asymptotic behavior of the
persistence and the frozen domain configurations on fractals
becomes manifest in Fig. 3(a), which shows the saturation value of
the persistence, $P_\infty$, as well as the fractions
$F_\infty(0)$, $F_\infty(1)$ and $F_\infty(2)$, as functions of
the coupling parameter for the GSG embedded in
Euclidean dimension $d=3$. Several discontinuities are observed in
the curve of $P_\infty$ in Fig. 3(a). The first discontinuity of
$P_\infty$ occurs at the value of the coupling $\epsilon_o=0.24$
where the fraction $F_\infty(0)$ vanishes. This implies that local
blocked configurations where a site has no neighbors in its same
phase, as those indicated in Fig. 2(a), become unstable at the
value of coupling $\epsilon_o$. The second discontinuity of $P_\infty$ takes
place at the value of coupling $\epsilon_1=0.30$ where $F_\infty(1)$ becomes
zero, and it is related to the loss of stability of local frozen
domain configurations as those signaled in Fig. 2(b). In addition,
$P_\infty$ reaches a minimum at the value of coupling
$\epsilon_2=0.36$, where the fraction $F_\infty(2)$ decays to zero. 
For  $\epsilon > \epsilon_2$ the domains of the two phases grow in
size as seen in Fig. 2(c) and 2(d). The phase domains also form faster,
reducing the number of phase switching of the elements and therefore
producing an increment in the saturation values $P_\infty$ up to
a maximum occurring at $\epsilon_{max}=0.72$.
\begin{figure}
\centerline{\hbox{
\epsfig{file=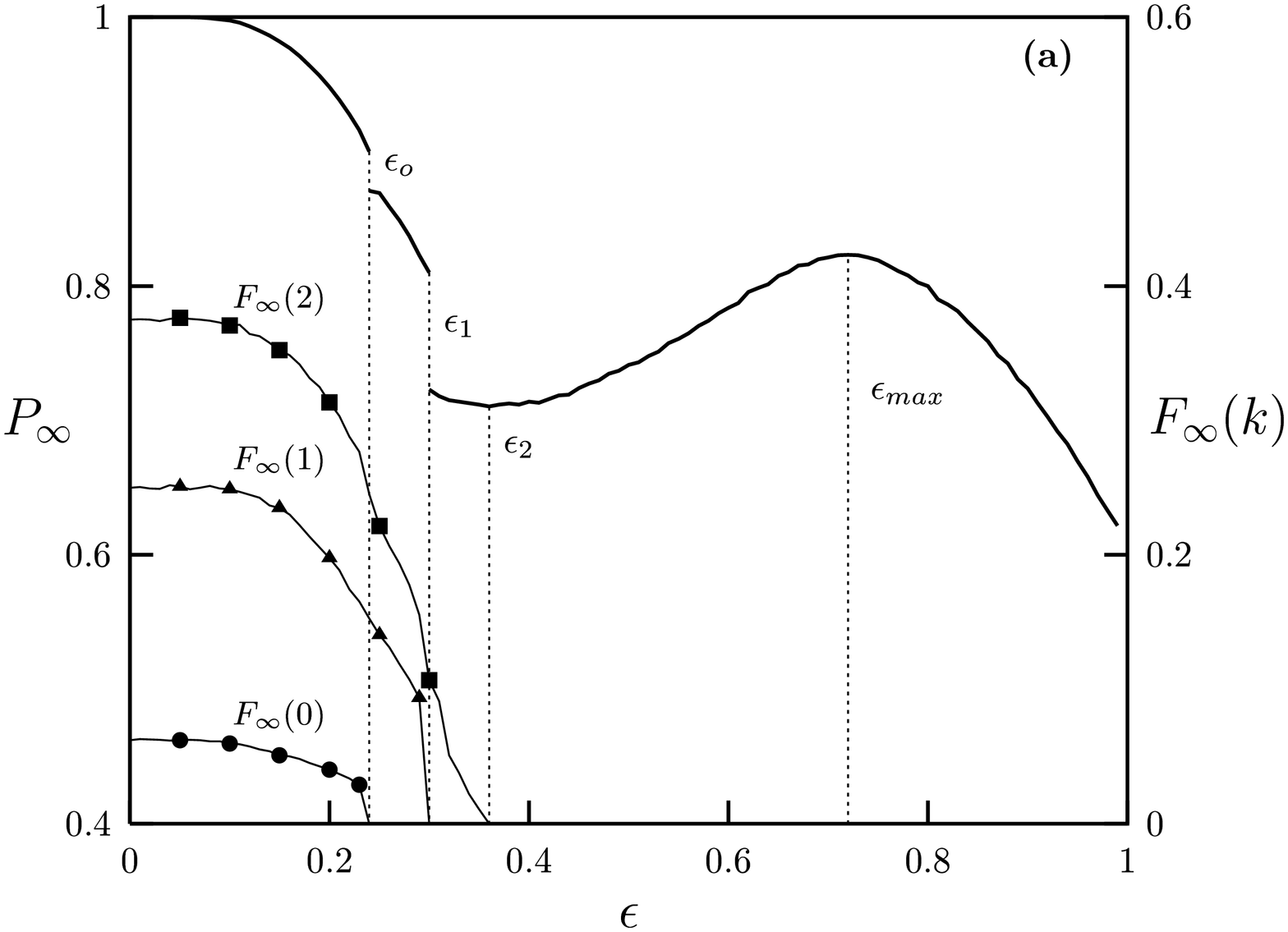,width=0.35\textwidth,clip=,angle=90} \hspace{-5mm}
}}
\centerline{\hbox{
\epsfig{file=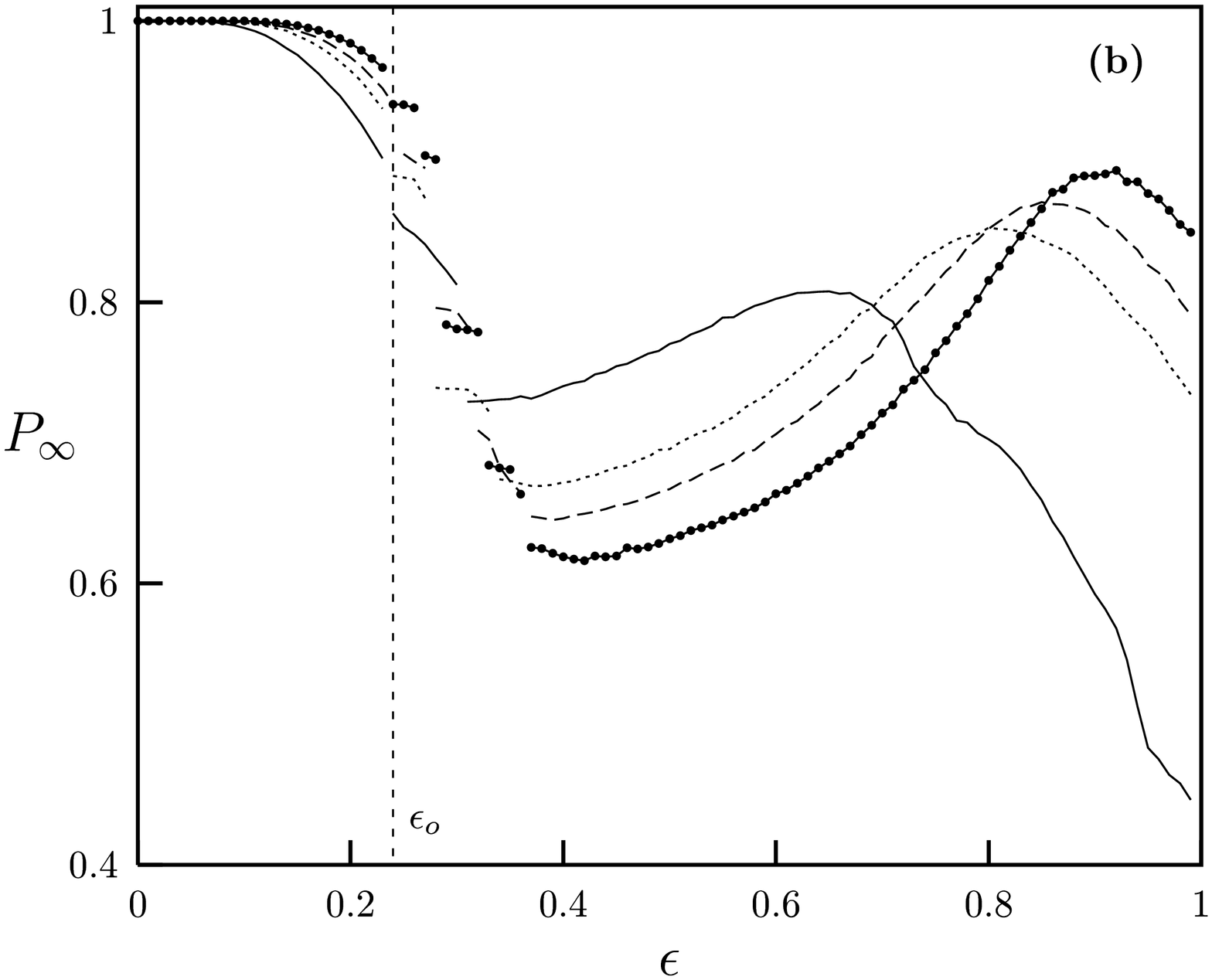,width=0.35\textwidth,clip=,angle=90}
}}
\caption{(a) Left scale: Saturation persistence $P_\infty$ as a function of
$\epsilon$ for the GSG embedded in Euclidean dimension $d=3$,
represented by the thick dark line. The values $\epsilon_o, \epsilon_1,
\epsilon_2$, and $\epsilon_{max}$  are indicated by dotted
vertical lines. Right scale: Fraction $F_t(k)$ vs. $\epsilon$.
Circles $F_\infty(0)$; triangles $F_\infty(1)$; squares
$F_\infty(2)$. Lattice size is $N=4^9$. 
(b) $P_\infty$ as a function of $\epsilon$ for GSG
embedded in different Euclidean dimensions $d$. Continuous
line: $d=2$, $N=3^{11}$; dotted line line: $d=5$, $N=6^7$; 
dashed line: $d=7$, $N=8^6$; circle line: $d=11$, $N=12^5$. The first
discontinuity at $\epsilon_o=0.24$, common to all lattices, is
indicated.} 
\end{figure}

A similar behavior is observed for
all the fractal networks embedded in different Euclidean dimensions $d$.
Figure~3(b) shows the saturation persistence $P_\infty$ as a function of
$\epsilon$ for GSG  associated to different $d$.
The discontinuities on each curve
are related to the loss of stability of the configurations where one local
element has a majority of its neighbors in the opposite phase. The number of
those configurations  is $J=\mbox{nint}((d+1)/2)$, where $\mbox{nint}(x)$ rounds
$x$ to its nearest integer.
These $J$ discontinuities take place at increasing values of
the coupling $\epsilon_k$ for which
the asymptotic fractions $F_\infty(k)$ vanish, with $k < J$.
The first discontinuity
is related to the vanishing of the fraction $F_\infty(0)$ at
the value $\epsilon_o=0.24$, independently of the embedding dimension $d$.
This independence is due to the presence of
the normalization factor
$(d+1)^{-1}$ in the coupling term in Eq.~(\ref{ec1}) for all the lattices.
For embedding dimensions $d$ odd, there occur a minimum of $P_\infty$ when
the fraction $F_\infty(J)$ becomes zero. In those cases 
the local configurations losing stability are those consisting of
a site with half of its neighbors in one phase and
the other half in the opposite phase. These local configurations are symmetric, 
in the sense that a change of phase of that site does not alter the phase
composition of its neighborhood.
These symmetric configurations can not
happen in GSG associated to $d$ even, and for those networks the minimum of
$P_\infty$ coincides with the the last discontinuity that takes place 
when $F_\infty(J-1)$ vanishes.

Note that for each fractal lattice there is a value of the coupling
$\epsilon_{max}$ where a maximum of $P_\infty$ is observed.
The origin of such maximum can be analyzed through
the average fraction of neighboring pairs that
have opposite phases at time $t$, defined as
\begin{equation}
G_t= 1-\frac{1}{N(d+1)} \sum_{i=1}^N   \left( \sum_{j \in {\cal
N}_i} \delta \left(\sigma_t(i),\sigma_t(j)\right) \right),
\end{equation}
where $\delta(\sigma_t(i),\sigma_t(j))=1$ if $\sigma_t(i)=\sigma_t(j)$, and
$\delta(\sigma_t(i),\sigma_t(j))=0$ otherwise.
In Fig. 4 we show $G_1$ as a function of the parameter $\epsilon$ for
different fractal lattices. The maximum of $G_1$ for each lattice 
takes place at the value $\epsilon_{max}$ at which the corresponding
curve of $P_\infty$ reaches a maximum. 
When $G_1$ is maximum the probability that there exist 
cells in one phase having all of their neighbors
in the opposite phase at the first iteration is also maximum. Therefore, for
the value of coupling $\epsilon_{max}$ there is a greater chance that 
at the next iteration
such cells change phase instead of their neighbors, yielding more 
stable domains with fewer
elements of the network having to switch their initial phase. Consequently,
the persistence probability, that measures the number of elements that have not
changed phases, is maximum at that time. 
Since the domains that are being formed
are the most stable, in successive times the persistence will sustain a
maximum value for the value of coupling $\epsilon_{max}$ corresponding to 
each lattice. For $\epsilon > \epsilon_{max}$ the coupling is strong enough
to induce transient changes in the phase of elements having 
the majority of their neighbors in that same phase and therefore 
producing lower values of $P_\infty$.

The local effect captured by the quantity $G_1$ in fractal
lattices also appears in
regular Euclidean lattices, although the asymptotic behavior of the persistence
is different in those two network topologies.
Figure 4 includes the calculation of $G_1$ as a function of
$\epsilon$ for a two-dimensional regular lattice; the maximum of $G_1$
in this case occurs at $\epsilon_{max}=0.67$. This is the critical value of
the coupling parameter found in Ref.~\cite{Chate}, after proper normalization,
for the phase ordering transition 
in the scaling behavior of the persistence in a two dimensional
Euclidean lattice. At $\epsilon=0.67$ the blocked states in the
two-dimensional regular lattice give place to growing domains of the two 
phases, separated by a continuous interface.
The interface motion is driven by curvature effects that cause changes of 
phase in many elements of the system and therefore a temporal decay in 
the persistence probability \cite{Chate2}.
\begin{figure}
\centerline{\hbox{
\epsfig{file=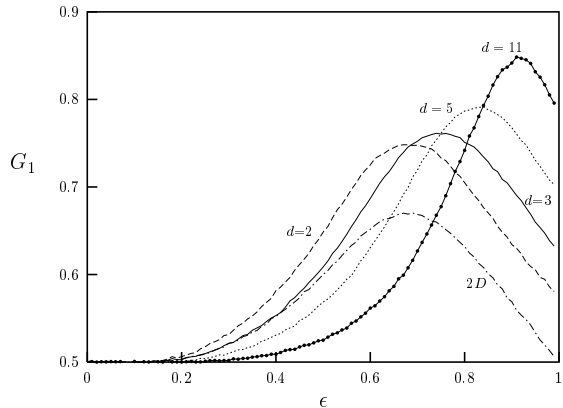,width=0.35\textwidth,clip=,angle=90}
}}
\caption{$G_1$ as a function of $\epsilon$ for
GSG embedded in different $d$.  Dashed line: $d=2$; continuous line: $d=3$;
dotted line: $d=5$; circle line: $d=11$. Sizes are the same as in
Fig.~3. Dotted-dashed line represents $G_1$ for a
two-dimensional regular lattice, $(2D)$.}
\end{figure}
In general, in regular Euclidean lattices the ratio between
the length of the interface to the size of a domain decays as
$r^{-1}$, where $r$ is the average radius of the domain.
On the other hand, in the fractal networks the interface consists of a few 
disconnected cells separating large domains and no curvature can be defined.
Because of the self-similarity of the structure the number of elements in a
domain grows as $r^{d_f}=(d+1)^l$, where $l$ is an integer, while the size
of the interface is of the order of $(d+1)$, independently of the
domain sizes. Thus the ratio of the interface to domain size
decreases as $r^{-d_f(l-1)/l}$ in the fractals.  
This decay is much faster than in regular Euclidean lattices and 
accelerates with increasing domain size, forming stable separated 
phase domains.
As a consequence, domains always freeze
on the fractal networks and the phase transition observed in the temporal
behavior of the persistence in regular Euclidean lattices
does not occur in fractals.

In summary, we have found that phase domains in  
chaotic maps coupled on fractal networks always reach a frozen configuration, 
causing the saturation of the persistence in time for all values of the 
coupling parameter, in contrast to the phase-transition behavior of the 
persistence observed in Euclidean regular lattices. 
The fractal nature of the spatial support is also reflected in the 
discontinuities observed in the $P_\infty$ vs. $\epsilon$
curves in Fig. 3. 
The phase configurations of the local neighborhoods have similar transient 
manifestations in fractal networks and in Euclidean regular lattices, 
as seen in the emergence of a maximum of $G_1$ at a value of coupling 
$\epsilon_{max}$. However, the asymptotic and global properties of the
phase ordering process on these 
two network topologies are quite different,
even when the number of local connections in the neighborhood
is the same, as it happens for the two-dimensional Euclidean
lattice and the GSG embedded in $d=3$. The necessary correlations
between elements for building complex collective dynamics are
more likely to occur in the denser Euclidean lattices.
These results suggest that the topology of the network and not the
number of local connections or the dimensionality of the space determine
the asymptotic collective behaviors that may emerge on 
networks of coupled chaotic elements.

This work was supported by Consejo de Desarrollo Cient\'{\i}fico,
Human\'{\i}stico y Tecnol\'ogico, Universidad de Los Andes,
M\'erida, Venezuela.


\begin{thebibliography}{99}
\bibitem{Chaos} Chaos {\bf 2}, No. 3 (1992), focus issue on
Coupled Map Lattices; edited by K. Kaneko.
\bibitem{CK2} M. G. Cosenza and R. Kapral, Chaos {\bf 4}, 99 (1994).
\bibitem{Gade} P. M. Gade, H. A Cerdeira, and R. Ramaswamy, Phys. Rev. E
{\bf 52}, 2478 (1995).
\bibitem{Tree} M. G. Cosenza and K. Tucci, Phys. Rev. E {\bf 64}, 026208 (2001).
\bibitem{Vol} D. Volchenkov, S. Sequeira, Ph. Blanchard, and M. G. Cosenza,
Stochastics and Dynamics {\bf 2}, 203 (2002).
\bibitem{Kay} M. G. Cosenza and K. Tucci, Phys. Rev. E {\bf 65}, 036223 (2002).
\bibitem{Amri} S. Jalan and R. E. Amritkar, Phys. Rev. Lett. {\bf 90}, 014101 (2003).
\bibitem{Chate} A. Lemaitre and H. Chat\'e, Phys. Rev. Lett. {\bf 82},
1140 (1999).
\bibitem{Chate2} J. Kockelkoren, A. Lemaitre, and H. Chat\'e, Physica A 
{\bf 288}, 326 (2000).
\bibitem{Wei} W. Wang, Z. Liu, and B. Hu, Phys. Rev. Lett. {\bf 84}, 2610 (2000).
\bibitem{Stra} L. Angelini, M. Pellicoro, ans S. Stramaglia, Phys. Lett. A
{\bf 285}, 293 (2001).
\bibitem{Just} F. Schm\"user, W. Just, and H. Kantz, Phys. Rev. E {\bf 61}, 
3675 (2000).
\bibitem{Derrida} B. Derrida, A. J. Bray, and C. Godreche, J. Phys. A
{\bf 27}, 357 (1994).
\end{thebibliography}
\end{document}